\documentstyle{aipproc}

\begin{document}
\title{Models of Neutrino Mass and Mixing}

\author{Ernest Ma}
\address{Department of Physics, University of California, Riverside, 
California, 92521}

\maketitle

\begin{abstract}
There are two basic theoretical approaches to obtaining neutrino mass and 
mixing.  In the minimalist approach, one adds just enough new stuff to the 
Minimal Standard Model to get $m_\nu \neq 0$ and $U_{\alpha i} \neq 1$. 
In the holistic approach, one uses a general framework or principle to 
enlarge the Minimal Standard Model such that, among other things, 
$m_\nu \neq 0$ and $U_{\alpha i} \neq 1$.  In both cases, there are 
important side effects besides neutrino oscillations.  I discuss a 
number of examples, including the possibility of leptogenesis from 
$R$ parity nonconservation in supersymmetry.
\end{abstract}

\section*{Introduction}

There are numerous studies of the neutrino mass matrix for explaining the 
disappearance and appearance of $\nu_e$ and $\nu_\mu$ in various 
experiments\cite{1}.  In this talk, I will only address the theoretical 
issue of how neutrinos obtain mass and the accompanying consequences 
beyond neutrino oscillations.  The starting point of any such discussion 
is the 1979 observation by Weinberg\cite{2} that given the particle content 
of the Minimal Standard Model at low energies, neutrinos acquire mass 
only through the following unique effective dimension-5 operator:
\begin{equation}
\Lambda^{-1} (\nu \phi^0 - e \phi^+)^2.
\end{equation}
A nonzero Majorana mass for $\nu$ is obtained as $\phi^0$ acquires a vacuum 
expectation value in electroweak symmetry breaking. Models of 
neutrino mass differ only in how this operator is realized\cite{3}.

\section*{Examples of the Minimalist Approach}

\noindent (1) Canonical Seesaw\cite{4}

Add 3 heavy singlet right-handed neutrinos to the Minimal Standard Model: 
1 $\nu_R$ for each $\nu_L$.  Then the Weinberg operator is realized 
because each heavy $\nu_R$ is linked to $\nu_L \phi^0$ with a Yukawa 
coupling $f$; and since $\nu_R$ is allowed to have a large Majorana mass 
$M_R$, the famous seesaw realtionship $m_\nu = m_D^2/M_R$ is obtained, 
where $m_D = f \langle \phi^0 \rangle$.  This mechanism dominates the 
literature and is usually implied when a particular pattern of neutrino 
mass and mixing is proposed.\\

\noindent (2) Minimal Seesaw\cite{5}

Add just 1 $\nu_R$.  Then only 1 linear combination of $\nu_e, \nu_\mu, 
\nu_\tau$ gets a seesaw mass.  The other 2 neutrino masses are zero at 
tree level, but since there is in general no more symmetry to protect 
their masslessness, they must become massive through radiative corrections. 
As it turns out, this happens in two loops through double $W$ exchange 
and the result is doubly suppressed by the charged-lepton masses.  Hence 
it is not a realistic representation of the present data for neutrino 
oscillations.\\

\noindent (3) Next-to-Minimal Seesaw\cite{6}

Add 1 $\nu_R$ and 1 extra Higgs doublet.  Then 1 neutrino gets a seesaw 
mass.  Another gets a one-loop mass through its coupling to $\phi_2^0$, 
where $\langle \phi_2^0 \rangle = 0$.  This second mass is proportional 
to the coupling of the term $(\bar \phi_2^0 \phi_1^0)^2$ times $\langle 
\phi_1^0 \rangle^2$ divided by $M_R$.  The third neutrino gets a two-loop 
mass as in (2).  This scheme is able to fit the present data.\\

\noindent (4) Purely Radiative Mechanism\cite{7}

Add 1 extra Higgs doublet $\Phi_2$ and 1 charged singlet $\chi^+$.  Then 
the coexistence of the terms $(\nu_i l_j - \nu_j l_i) \chi^+$ and $(\phi_1^+ 
\phi_2^0 - \phi_2^+ \phi_1^0) \chi^-$ allows the following radiative mass 
matrix to be obtained:
\begin{equation}
{\cal M}_\nu = \left[ \begin{array}{c@{\quad}c@{\quad}c} 0 & f_{\mu e} 
(m_\mu^2-m_e^2) & f_{\tau e} (m_\tau^2-m_e^2) \\ f_{\mu e} (m_\mu^2-m_e^2) 
& 0 & f_{\tau \mu} (m_\tau^2-m_\mu^2) \\ f_{\tau e} (m_\tau^2-m_e^2) & 
f_{\tau \mu} (m_\tau^2-m_\mu^2) & 0 \end{array} \right].
\end{equation}
This model has been revived in recent years and may be used to fit the 
neutrino-oscillation data.\\

\noindent (5) Heavy Higgs Triplet\cite{8}

Add 1 heavy Higgs triplet $(\xi^{++}, \xi^+, \xi^0)$.  Then the coexistence 
of the terms $\nu_i \nu_j \xi^0$ and $\phi^0 \phi^0 \bar \xi^0$ allows 
a tree-level neutrino mass given by
\begin{equation}
m_\nu = {2 f \mu \langle \phi^0 \rangle^2 \over m_\xi^2} = 2 f \langle 
\xi^0 \rangle.
\end{equation}
This shows the interesting result that $\xi$ has a very small vacuum 
expectation value inversely proportional to the square of its mass. 
Note also that the effective operator of Eq.~(1) should now be written as
\begin{equation}
(\nu_i \phi^0 - l_i \phi^+)(\nu_j \phi^0 - l_j \phi^+) = 
 \nu_i \nu_j (\phi^0 \phi^0) - (\nu_i l_j + l_i \nu_j)(\phi^0 \phi^+) + 
l_i l_j (\phi^+ \phi^+),
\end{equation}
which shows clearly the role of $\xi$.\\

\noindent (6) Radiative Splitting of Neutrino Mass Degeneracy\cite{9}

Add 1 Higgs triplet as in (5).  Assume further that
\begin{equation}
{\cal M}_\nu = \left[ \begin{array} {c@{\quad}c@{\quad}c} 0 & m_0 & 0 \\ 
m_0 & 0 & 0 \\ 0 & 0 & m_3 \end{array} \right]
\end{equation}
at tree level.  Then the mass eigenstates corresponding to the mass 
eigenvalues $\pm m_0$ are radiatively corrected to have slightly different 
masses at one-loop level, resulting in the following successful connection 
between atmospheric and solar neutrino vacuum oscillations:
\begin{equation}
{(\Delta m^2)_{sol} (\Delta m^2)_{atm} \over m_\nu^4 (\sin^2 2 \theta)_{atm}} 
= 2 I^2,
\end{equation}
where $I = (3 G_F m_\tau^2/16 \pi^2 \sqrt 2) \ln (m_\xi^2/m_W^2)$, and
\begin{equation}
\left( \begin{array} {c} \nu_1 \\ \nu_2 \\ \nu_3 \end{array} \right) = 
\left( \begin{array} {c@{\quad}c@{\quad}c} 1/\sqrt 2 & -c/\sqrt 2 & s/\sqrt 2 
\\ 1/\sqrt 2 & c/\sqrt 2 & -s/\sqrt 2 \\ 0 & s & c \end{array} \right) 
\left( \begin{array} {c} \nu_e \\ \nu_\mu \\ \nu_\tau \end{array} \right)
\end{equation}
has been assumed.

\section*{Some Generic Consequences}

\noindent (A) Once neutrinos have mass and mix with one another, the 
radiative decay $\nu_2 \to \nu_1 \gamma$ happens in all models, but is 
usually harmless as long as $m_\nu <$ few eV, in which case it will have 
an extremely long lifetime, many many orders of magnitude greater than the 
age of the Universe.\\

\noindent (B) The analogous radiative decay $\mu \to e \gamma$ also happens 
in all models, but is only a constraint for some models where $m_\nu$ is 
radiative in origin, such as in (3).\\

\noindent (C) Neutrinoless double $\beta$ decay occurs\cite{10}, but is 
sensitive only to the $\nu_e - \nu_e$ entry of ${\cal M}_\nu$, which may 
be assumed to be zero as in (6).\\

\noindent (D) Leptogenesis is possible in the 2 simplest models of neutrino 
mass, {\it i.e.} (1) and (5).  In the canonical seesaw scenario, $\nu_R$ 
may decay into both $l^- \phi^+$ and $l^+ \phi^-$\cite{11}.  In the Higgs 
triplet scenario, $\xi^{++}$ may decay into both $l^+ l^+$ and $\phi^+ 
\phi^+$\cite{8}.  The lepton asymmetry thus generated may be converted into 
the present observed baryon asymmetry of the Universe through the 
electroweak sphalerons\cite{12}.

\section*{Examples of the Holistic Approach}

\noindent (7) Grand unification usually requires new particles at high 
energies and lepton-number conservation to be violated at some scale. 
Hence it is ideal for the consideration of neutrino mass.  There is a 
vast literature on this subject and I will not discuss anything more in 
this talk other than the simple observations that $SO(10)$ contains 
$\nu_R$ and that $E_6$ contains both $\nu_R$ and $\nu_S$, where the latter 
may be regarded as a sterile neutrino which has a natural reason to be 
light\cite{13}.\\

\noindent (8) $R$ parity nonconserving supersymmetry is another 
very fruitful approach which has received a lot of attention in the past 
2 years or so.  If only $B$ is assumed to be conserved but not $L$, then 
the superpotential also contains the terms
\begin{equation}
\mu_i L_i H_2 + \lambda_{ijk} L_i L_j e_k^c + \lambda'_{ijk} L_i Q_j d_k^c,
\end{equation}
which violates $R \equiv (-1)^{3B+L+2J}$.  As a result, a radiative 
neutrino mass $m_\nu \simeq \lambda'^2 (A m_b^2)/16 \pi^2 m_{\tilde b}^2$ 
may be obtained\cite{14}.  Furthermore, from the mixing of $\nu_i$ 
with the neutralino mass matrix through the bilinear term $L_i H_2$ and 
the induced vacuum expectation value of $\tilde \nu_i$, a tree-level 
mass $m_\nu \simeq (\mu_i/\mu - \langle \tilde \nu_i \rangle /\langle 
h_1^0 \rangle )^2 m_{eff}$ is also obtained\cite{15}.

\section*{More Side Effects}

\noindent (E) New particles at the 100 GeV mass scale exist in some 
radiative models.  They can be searched for in future accelerators.\\

\noindent (F) Lepton-flavor changing processes at tree level provide 
another mechanism for matter-induced neutrino oscillations.\\

\noindent (G) Lepton-number violating interactions at the TeV mass scale 
may erase any preexisting $B$ or $L$ asymmetry of the Universe\cite{16}. 
In $R$ parity nonconserving supersymmetry, $\lambda' > 10^{-4}$ is required 
for realistic $m_\nu$, but $\lambda' < 10^{-7}$ is needed to avoid 
erasure\cite{17}.

\section*{Leptogenesis from R Parity Violation}

As remarked already earlier in (G), whereas lepton-number violating trilinear 
couplings in Eq.~(8) are able to generate neutrino masses radiatively, they 
also wash out any preexisting $B$ or $L$ asymmetry during the electroweak 
phase transition.  On the other hand, successful leptogenesis may still be 
possible as shown recently\cite{18}.

Assume the lightest and 2nd lightest supersymmetric particles to be
\begin{equation}
\tilde W'_3 = \tilde W_3 - \epsilon \tilde B, ~~~ \tilde B' = \tilde B + 
\epsilon \tilde W_3,
\end{equation}
respectively, where $\tilde W_3$ and $\tilde B$ are the $SU(2)$ and $U(1)$ 
neutral gauginos, and $\epsilon$ is a very small number.  Note that 
$\tilde B$ couples to $\bar \tau_L^c \tilde \tau_L^c$ but $\tilde W_3$ 
does not, because $\tau_L^c$ is trivial under $SU(2)$.  Assume $\tilde \tau_L 
- h^-$ mixing to be negligible but $\tilde \tau_L^c - h^+$ mixing to be 
significant and denoted by $\xi$.  Obviously, $\tilde \tau$ may be repalced 
by $\tilde \mu$ or $\tilde e$ in this discussion.

Given the above assumptions, $\tilde B'$ decays into $\tau^\mp h^\pm$ 
through $\xi$, whereas $\tilde W'_3$ decays (also into $\tau^\mp h^\pm$) 
are further suppressed by $\epsilon$.  This allows $\tilde W'_3$ decay to 
be slow enough to be out of equilibrium with the expansion of the Universe 
at a temperature $\sim$ 2 TeV, and yet have a large enough asymmetry 
$(\tau^- h^+ - \tau^+ h^-)$ in its decay to obtain $n_B/n_\gamma \sim 
10^{-10}$.  See Figure 1.

This unique scenario requires $\tilde W'_3$ to be lighter than $\tilde B'$ 
and that both be a few TeV in mass so that the electroweak sphalerons are 
still very effective in converting the $L$ asymmetry into a $B$ asymmetry.  
It also requires very small mixing bewteen $\tilde \tau_L$ with $h^-$, 
which is consistent with the smallness of the neutrino mass required in 
the phenomenology of neutrino oscillations.  On the other hand, the mixing 
of $\tilde \tau_L^c$ with $h^+$, i.e. $\xi$, should be of order $10^{-3}$ 
which is too large to be consistent with the usual terms of soft 
supersymmetry breaking.  For successful leptogenesis, 
the nonholomorphic term $H_2^\dagger H_1 \tilde \tau_L^c$ is required.

\section*{CONCLUSION}

Models of neutrino mass and mixing invariably lead to other possible physical 
consequences which are important for our overall understanding of the 
Universe, as well as other possible experimentally verifiable predictions.

\section*{ACKNOWLEDGEMENTS}

I thank D. Cline and the other organizers of $\mu \mu 99$ for a great 
meeting.  This work was supported in part by the U.~S.~Department of Energy 
under Grant No. DE-FG03-94ER40837.

\begin{figure}[htb]
\mbox{}
\vskip 3.8in\relax\noindent\hskip .5in\relax
\includegraphics{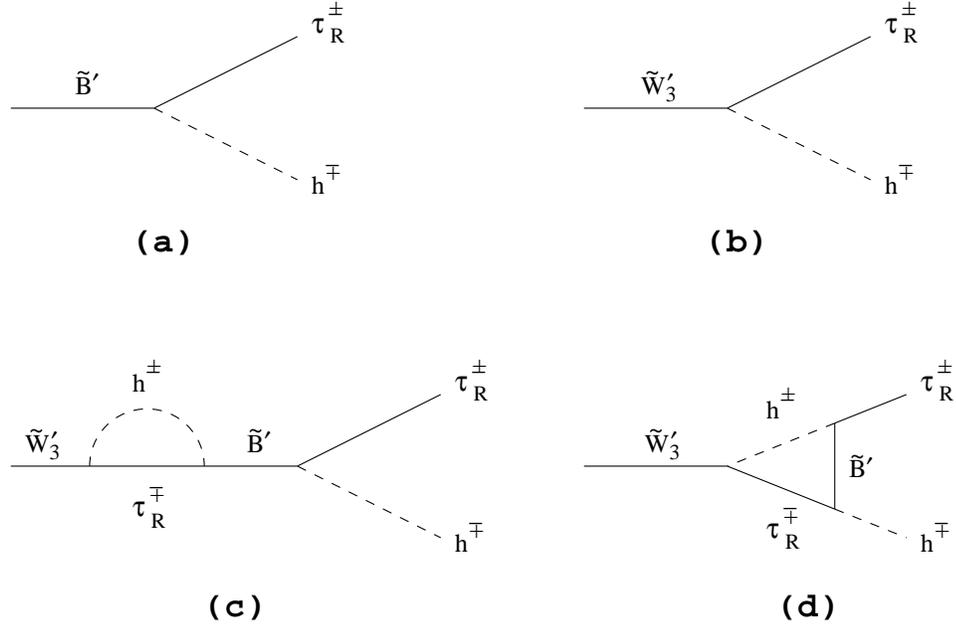} 
\vskip .25in
\caption{ Tree-level diagrams for (a) $\tilde B'$ 
decay and (b) $\tilde W'_3$ decay (through their $\tilde B$ content), 
and the one-loop (c) self-energy and (d) vertex diagrams for  
$\tilde W'_3$ decay which have absorptive parts of opposite lepton number.}
\end{figure}

\end{document}